\documentclass[a4paper,twocolumn,english,aps, prl]{revtex4}
\usepackage[T1]{fontenc}
\usepackage[latin9]{inputenc}
\usepackage{graphicx}
\usepackage{amssymb}

\usepackage{parskip}
\makeatletter

\pdfpageheight\paperheight
\pdfpagewidth\paperwidth

\DeclareRobustCommand{\greektext}{%
  \fontencoding{LGR}\selectfont\def\encodingdefault{LGR}}
\DeclareRobustCommand{\textgreek}[1]{\leavevmode{\greektext #1}}
\DeclareFontEncoding{LGR}{}{}
\DeclareTextSymbol{\~}{LGR}{126}

\@ifundefined{textcolor}{}
{%
 \definecolor{BLACK}{gray}{0}
 \definecolor{WHITE}{gray}{1}
 \definecolor{RED}{rgb}{1,0,0}
 \definecolor{GREEN}{rgb}{0,1,0}
 \definecolor{BLUE}{rgb}{0,0,1}
 \definecolor{CYAN}{cmyk}{1,0,0,0}
 \definecolor{MAGENTA}{cmyk}{0,1,0,0}
 \definecolor{YELLOW}{cmyk}{0,0,1,0}
 }

\makeatother

\usepackage{babel}

\begin{document}

\title{Finding All the Stationary Points of a Potential Energy Landscape
via Numerical Polynomial Homotopy Continuation Method}

\author{Dhagash Mehta}
\email{dbmehta@syr.edu}

\affiliation{Department of Physics, Syracuse University, Syracuse, NY 13244, USA}
\begin{abstract}
The stationary points (SPs) of a potential energy landscape play a
crucial role in understanding many of the physical or chemical properties
of a given system. Unless they are found analytically,
there is, however, no efficient method to obtain \textit{all }the
SPs of a given potential. We introduce a novel method, called the
numerical polynomial homotopy continuation (NPHC) method, which numerically
finds all the SPs, and is \textit{embarrassingly
parallelizable}. The method requires the non-linearity of the potential
to be polynomial-like, which is the case for almost all of the potentials
arising in physical and chemical systems. We also \textit{certify}
the numerically obtained SPs so that they are independent
of the numerical tolerance used during the computation. It is then
straightforward to separate out the local and global minima. As a
first application, we take the XY model with power-law interaction
which is shown to have a polynomial-like non-linearity and apply
the method.
\end{abstract}
\maketitle
\noindent \textbf{Introduction:} A Potential energy landscape (PEL)
is the hyper-surface of some given potential $V(\vec{x})$, with $\vec{x}=(x_{1},x_{2},...,x_{N})$
being the variables (e.g., position coordinates, fields etc.). Studying
the stationary points (SPs), defined by the solutions of
the system of $N$ equations $\frac{\partial V(\overrightarrow{x})}{\partial x_{i}}=0,i=1,...,N$, of the PEL
is crucial in learning many physical or chemical properties of the
system described by $V(\vec{x})$. The SPs are classified according to the number of negative
eigenvalues of the Hessian matrix $\mathcal{H}$ evaluated at each
SP: the SPs with no negative eigenvalue are called minima and the
SPs with at least one negative eigenvalue are called saddles. In statistical
mechanics, for example, the stationary points of the PEL have been
shown to be directly related to the non-analyticity of thermodynamic
quantities (i.e., phase transitions) in the respective models \cite{RevModPhys.80.167};
the global minimum of a spin glass model is invaluable for studying its equilibrium
properties. In theoretical chemistry, studying the properties
of the PEL of supercooled liquids and glasses has been a very active
area of research \cite{Wales:04,2011arXiv1104.1343C}, specifically,
in the study of Kramer's reaction rate theory for the thermally activated
escape from metastable states, and the computation of various physical
quantities like the diffusion constant using the minima of the PEL,
etc.~\cite{Wales:04}. In string theory, the SPs of the PEL of various
supersymmetric potentials correspond to the so-called string vacua. A lot of current activities in the string phenomenology areas have been focused
on developing different methods to find these string vacua \cite{Gray:2008zs,Gray:2006gn,Gray:2007yq}. 

If \textit{all} SPs are found analytically for the given $V(\overrightarrow{x})$,
then the problem is settled, obviously. But if the analytical solutions
are intractable, then one has to rely on alternative methods. Though
finding SPs is of the utmost importance in so many areas, there do
not exist many rigorous methods to find SPs, compared to the number
of methods for minimizing a potential.

One such method is the gradient-square minimization method in which
one minimizes $W=|\nabla V(\overrightarrow{x})|^{2}$ using some traditional
numerical minimization method such as Conjugate Gradient, Simulated
Annealing, etc. \cite{PhysRevLett.85.5356,PhysRevLett.85.5360}. The
minima of $W$, with further restriction that $W=0$, are the SPs
of $V(\overrightarrow{x})$.  However, there also exist minima of $W$ where 
$W>0$, and it has been shown that the number of these non-SPs grow as the system size 
increases \cite{2002JChPh.116.3777D,2003JChPh.11912409W}, thus making the method inefficient. 

The Newton-Raphson method (and its sophisticated variants) may also be used to
solve the system of $N$ non-linear equations $\frac{\partial V(\overrightarrow{x})}{\partial x_{i}}=0,i=1,...,N$.  
Here, an initial random guess is fed and then refined to a given numerical
precision to obtain a solution of the system \cite{2002PhRvL..88e5502G,2002JChPh.116.3777D}.
However, this method still suffers the major drawbacks of the gradient-square minimization method, 
namely, the possible existence of large basins of attraction may lead us to repeatedly obtain the same SPs and, 
more importantly, no matter how many times we iterate the respective algorithms with different initial
guesses, we are never sure if we have found \textit{all} SPs.

If the system of stationary equations has polynomial-like non-linearity,
then the situation is a bit better: in the string theory community,
for such a system of equations, the symbolic methods based on the
Gr\"{o}bner basis technique are used to solve the system \cite{Gray:2006gn,Gray:2008zs,Gray:2007yq}
which ensure that all the SPs are obtained when the computation finishes.
Roughly speaking, for a given system of multivariate polynomial equations
(which is known to have only isolated solutions), the so-called Buchberger
Algorithm (BA) or its refined variants can compute a new system of
equations, called a Gr\"{o}bner basis, in which the first equation only
consists of one of the variables and the subsequent equations consist
of increasing number of variables. The solutions of the new system
remain the same as the original system, but the former is easier to
solve. This method apparently resolves all of the above mentioned
problems, however, there are a few other problems: the BA is known
to have suffered from the exponential space complexity, i.e., the
memory (Random Access Memory) required by the machine blows up exponentially
with the number of variables, equations, terms in each polynomial,
etc.). So even for small sized systems, one may not be able to compute
a Gr\"{o}bner basis. It is also inefficient for the systems with irrational
coefficients. Furthermore, the BA is highly sequential.

In this Letter, we present the numerical polynomial homotopy continuation (NPHC) method which solves all the 
above problems for systems of equations with polynomial-like non-linearity. Numerical continuation methods 
have been around for some time \cite{79:allgower} and the polynomial homotopy continuation method has also been an 
active area of research \cite{SW:95, Li:2003} (see
\cite{Mehta:2009,2009iwqg.confE..25M} for the earlier account on the NPHC
method for various areas in theoretical physics). As with the Gr\"{o}bner
basis technique, the method extensively uses concepts from complex algebraic geometry and hence we allow the 
variables to take values from the complex space, even if the physically important solutions are
only the real ones. So long as a system of polynomial equations is known to have only isolated solutions, 
the NPHC guarantees that we obtain \textit{all} complex solutions of the system, from which all the real solutions 
can subsequently be filtered out.

\noindent \textbf{Numerical Polynomial Homotopy Continuation Method:}
Here we introduce the NPHC method to solve a system of multivariate
polynomial equations. Specifically, we consider a system $P(\vec{x})=0$ which is \textit{known to have isolated 
solutions} and where 
$P(\vec{x})=(\frac{\partial V(\vec{x})}{\partial x_{1}},\dots,\frac{\partial V(\vec{x})}{\partial x_{N}})$. 
Now, the \textit{Classical
B\'{e}zout Theorem} asserts that for a system of $N$ polynomial equations
in $N$ variables that is known to have only isolated solutions, the maximum
number of solutions in $\mathbb{C}^{N}$ is $\prod_{i=1}^{N}d_{i}$,
where $d_{i}$ is the degree of the $i$th polynomial. This bound
is called the classical B\'{e}zout bound (CBB).

Based on the CBB, a \textit{homotopy} can be constructed as $H(\vec{x},t)=\gamma(1-t)Q(\vec{x})+t\, P(\vec{x})=0$,
where $\gamma$ is a random complex number. The new system $Q(\vec{x})=(q_{1}(\vec{x}),\dots,q_{N}(\vec{x}))$,
called the \textit{start system}, is a system of polynomial equations
with the following properties: (1) the solutions of $Q(\vec{x})=H(\vec{x},0)=0$
are known or can be easily obtained. The solutions of $Q(\vec{x})=0$
are called start solutions; (2) the number of solutions of $Q(\vec{x})=H(\vec{x},0)=0$
is equal to the CBB of $P(\vec{x})=0$, (3) the solution set of $H(\vec{x},t)=0$
for $0\le t\le1$ consists of a finite number of smooth paths, each
parametrized by $t\in[0,1)$, and (4) every isolated solution of $H(\vec{x},1)=P(\vec{x})=0$
can be reached by some path originating at a solution of $H(\vec{x},0)=Q(\vec{x})=0$.
We can then track all the paths corresponding to each start solution
from $t=0$ to $t=1$ and reach (or diverge from) $P(\vec{x})=0=H(\vec{x},1)$.
It is rigorously shown that for a generic value of complex $\gamma$,
all paths are regular for $t\in[0,1)$, i.e., there is no singularity
along the path \cite{SW:95}. By implementing an efficient path tracker
algorithm, such as the Euler's predictor and Newton's corrector method,
all isolated solutions of $P(\vec{x})=0$ can be obtained. We do not
delve into the discussion of the actual path tracker algorithms used
in practice in this Letter, except mentioning that in the path tracker
algorithms used in practice, almost all apparent difficulties have
been resolved, such as tracking singular solutions, multiple roots,
solutions at infinity, etc.\cite{SW:95}.

As a trivial illustration, let us take the univariate polynomial, $P(x)=x^{2}-5=0$, from \cite{SW:95}. 
To find its solutions, we may for example choose $Q(x)=(x^{2}-1)$ as our start system as it satisfies the 
twin criteria that it has the same number of solutions as the CBB of $P(x)$ and is also easily solved to 
obtain the start solutions $x=\pm1$. (Note that, more generally, $Q(\vec{x})=(x_{1}^{d_{1}}-1,\dots,x_{N}^{d_{N}}-1)$
is the simplest choice as a start system for the multivariate case.) 
The problem of getting all solutions of $P(x)=0$ now reduces to simply tracking the solutions 
of $H(x,t)=0$ from $t=0$ to $t=1$ so that the paths beginning at $x=\pm1$ lead us to 
the actual solutions $x=\pm\sqrt{5}$.  Note that, in general, if there are more start solutions
than actual solutions, then the remaining paths diverge as $t$ approaches $1$.

There are several sophisticated computational packages such as PHCpack
\cite{Ver:99}, Bertini \cite{BHSW06} and
HOM4PS2 \cite{L:03} which can be used to solve systems of univariate
and, more importantly, multivariate polynomial equations, and are
available as freeware.  For the sake of completeness, the solutions of the above mentioned univariate equation
are $x=\pm(2.236067977+i\,10^{-10})$, i.e., $\pm\sqrt{5}$
up to the numerical precision. 

Since each path can be tracked independently of all others,
the NPHC is known as being \textit{embarrassingly parallelizable}, a feature 
that makes it very efficient.

For the multivariate case, a solution is a set of numerical values
of the variables which satisfies each of the equations with a given
tolerance, $\triangle_{\mbox{sol}}$ ($\sim10^{-10}$ in our set up).
Since the variables are allowed to take complex values, all the solutions
come with real and imaginary parts. A solution is a real solution
if the imaginary part of each of the variables is less than or equal
to a given tolerance, $\triangle_{\mathbb{R}}$ ($\sim10^{-7}$ is
a robust tolerance for the equations we will be dealing with in the
next section, below which the number of real solutions does not change).
All these solutions can be further refined with an \textit{arbitrary
precision} up to the machine precision. 

The obvious question at this stage would be if the number of real
solutions depend on $\triangle_{\mathbb{R}}$. To resolve this issue, we use a very recently developed
algorithm called alphaCertified which is based on the so-called \textgreek{a}-theory
to certify the real non-singular solutions of polynomial systems using
both exact rational arithmetic and arbitrary precision fl{}oating
point arithmetic \cite{2010arXiv1011.1091H}. This is a remarkable
step, because using alphaCertified we can prove that a solution classified
as a real solution is actually a real solution independent of $\triangle_{\mathbb{R}}$,
and hence these solutions are as good as the \textit{exact solutions}.

\noindent \textbf{A First Application:} As a first application, we
choose the XY model with long range power-law (algebraically decaying)
interaction, defined as $V(\vec{\theta})=K\sum_{i=1}^{N}\sum_{j=1}^{\frac{(N-1)}{2}}\frac{1-\cos(\theta_{i}-\theta_{i+j})}{j^{\alpha}},$
where $N$ is odd for our purposes, the normalization constant $K=(2\sum_{j=1}^{\frac{(N-1)}{2}}\frac{1}{j^{\alpha}})^{-1}$,
and $\alpha\in[0,\infty)$ \cite{2010arXiv1011.5050K}. $\alpha=0$ reproduces the mean-field
XY model and $\alpha\rightarrow\infty$ reproduces the nearest-neighbour
coupling XY model for which all the SPs are analytically obtained
recently \cite{Mehta:2009,vonSmekal:2007ns,Mehta2011}. We choose
$\alpha=0.75$ for which the coefficients take values from $\mathbb{R}$,
unlike for example $\alpha=1$ for which the integers are rational
numbers. Hence, we are already in the domain of problems where the
Gr\"{o}bner basis technique is inefficient. We impose periodic boundary
condition, i.e., $\theta_{k+N}=\theta_{k}$. Spin glass
models with power-law interaction have gained a huge interest recently
\cite{Katzgraber201035}. The chosen model is one of the simplest
models of this kind with a continuous symmetry. The model is also
known as the Kuramoto model with power-law interaction. The stationary
equations are $\frac{\partial V(\vec{\theta})}{\partial\theta_{k}}=K\sum_{j=1}^{\frac{(N-1)}{2}}(\frac{\sin(\theta_{k}-\theta_{k+j})+\sin(\theta_{k}-\theta_{k-j})}{j^{\alpha}})=0,$
for $k=1,\dots,N$. To get rid of the global rotation symmetry $\theta_{k}\rightarrow\theta_{k}+\phi$
where $\phi\in\mathbb{R}$, we fix one of the angles, say $\theta_{N}=0$,
and remove the $N$th equation out of the system, as done in \cite{Mehta2011,Mehta:2009}.
Kastner in \cite{2010arXiv1011.5050K} used a specific class of known
solutions, called the stationary wave solutions, i.e., $\theta_{m}^{(n)}=\frac{2\pi mn}{N}$,
where $m,n\in\{1,\dots,N$\}. Below we find that there are many more
solutions for this model.

The above system of equations is not apparently a system of polynomial
equations. But we can transform it into one \cite{Mehta:2009,2009iwqg.confE..25M}
by first using the trigonometric identities,
$\sin(\theta_{k}-\theta_{k+j})=\sin\theta_{k}\cos\theta_{k+j}-\sin\theta_{k+j}\cos\theta_{k}$
etc.; abbreviating each $\sin\theta_{k}=s_{k}$ and $\cos\theta_{k}=c_{k}$, in all the $N-1$ equations; and finally
adding $N-1$ additional constraint equations as $s_{k}^{2}+c_{k}^{2}-1=0$, for $k=1,\dots,N-1$, we get a system of $2(N-1)$ polynomial equations
consisting of $2(N-1)$ algebraic variables $c_{k}$s and $s_{k}$s as below:
\begin{eqnarray}
K\sum_{j=1}^{\frac{(N-1)}{2}}\frac{(s_{k}c_{k+j}-s_{k+j}c_{k}+s_{k}c_{k-j}-s_{k-j}c_{k})}{j^{\alpha}}=0,\nonumber \\
s_{k}^{2}+c_{k}^{2}-1=0,\label{eq:alg_dec_XY_model_poly}
\end{eqnarray}
for $k=1,\dots,N-1$. The removal of the global rotation symmetry
ensures that the system will have only isolated solutions. We
can now use the NPHC to solve the Eqs. (\ref{eq:alg_dec_XY_model_poly})
for all the $s_{k}s$ and $c_{k}$s; filter the real solutions out
using the above mentioned tolerances; and finally get the original
$\theta$-variables back by $\theta_{k}=\tan^{-1}(\frac{s_{k}}{c_{k}})\in(-\pi,\pi]$,
for all $k=1,\dots,N-1$. The results are as follows.

\begin{figure}
\includegraphics[width=4cm,height=3cm]{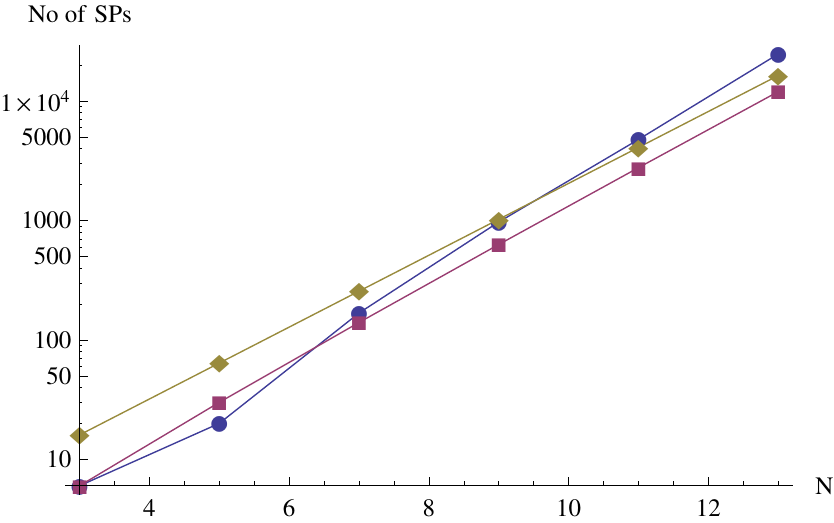}\includegraphics[width=4cm,height=3cm]{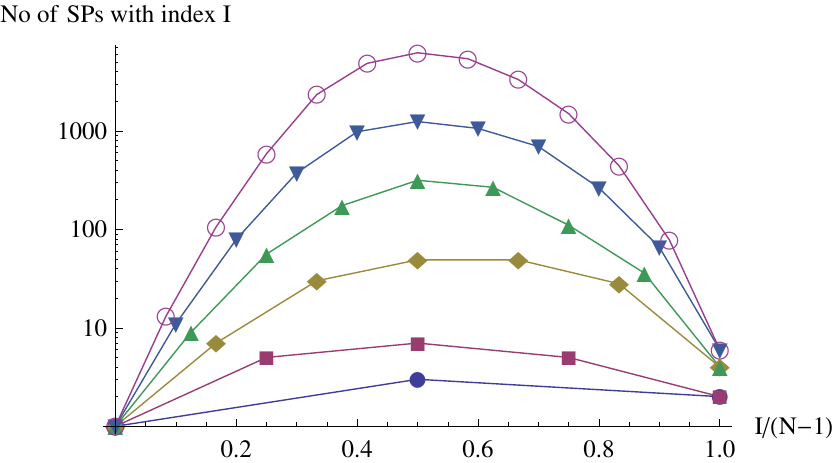}

\caption{The left panel is the plot for $N$ vs no. of SPs: the circles, squares
and diamonds represent data points for the algebraic decaying XY model,
the one-dimensional nearest-neighbour XY model for periodic \cite{Mehta2011,Mehta:2009}
and anti-periodic conditions \cite{vonSmekal:2007ns}, respectively.
The lines are drawn for a guide to the eyes. The right panel is the
plot for index $I$ vs no of SPs with index $I$. From top to bottom,
$N=13,11,9,7,5,3$. The normalization on $I$ is $N-1$ since we have
taken $\theta_{N}=0$. \label{fig:N_vs_SPs}}
\vspace{-0.5cm}
\end{figure}

\begin{figure}
\centering
\includegraphics[width=4cm,height=3cm]{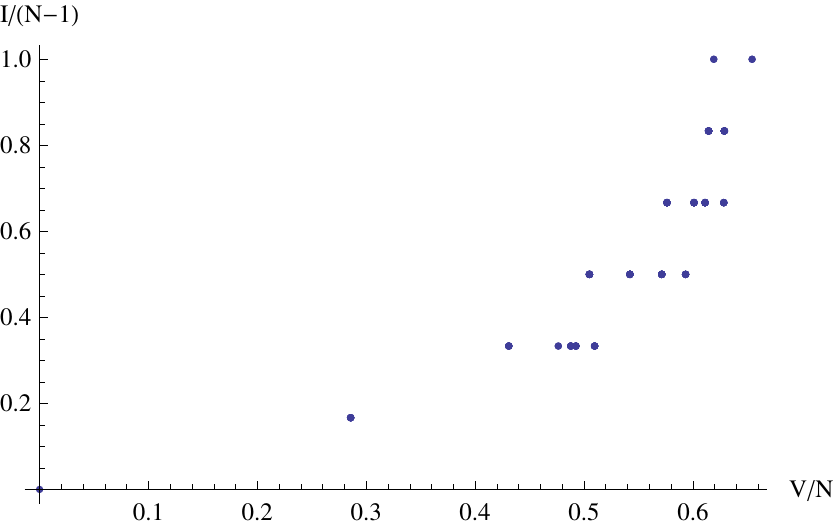}\includegraphics[width=4cm,height=3cm]{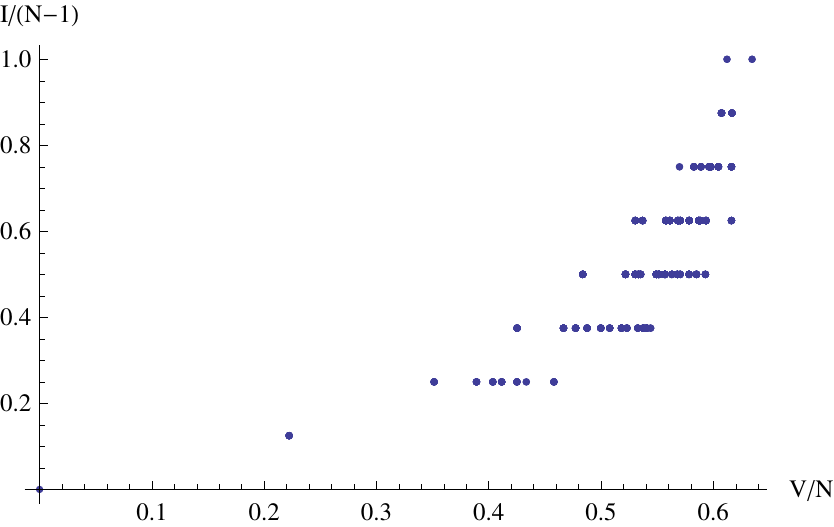}

\includegraphics[width=4cm,height=3cm]{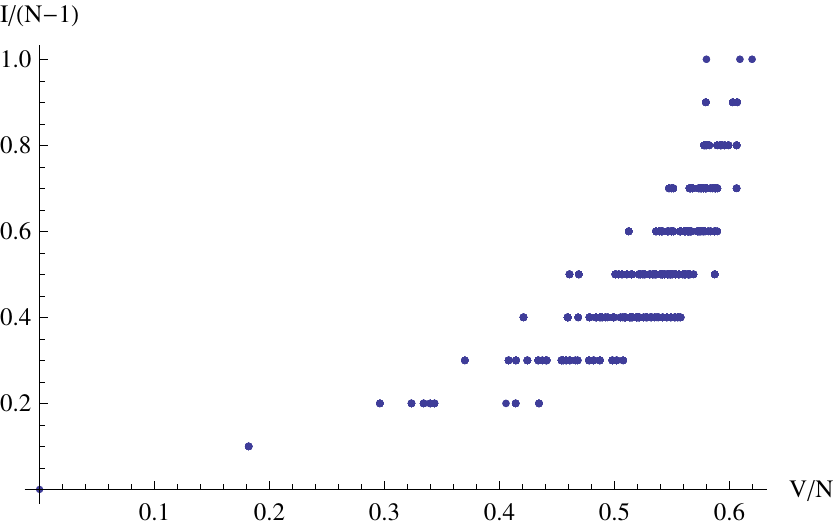}\includegraphics[width=4cm,height=3cm]{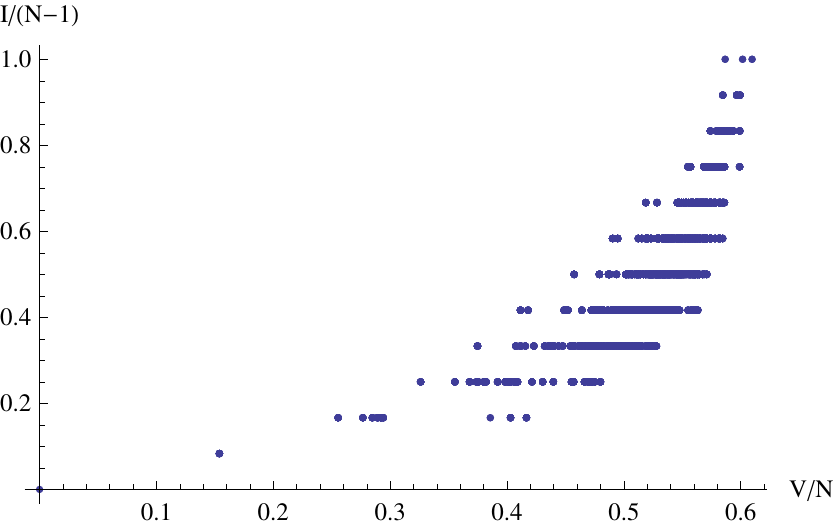}
\caption{Plot for $V/N$ vs index density, starting from top left corner $N=7,9,11,13$,
respectively.\label{fig:energy_density_vs_index} }
\vspace{-0.5cm}
\end{figure}

Firstly, the number of SPs for $N=3,5,7,9,11,13$ are $6,20,168,972,4774,24830$,
respectively. The left panel in Figure \ref{fig:N_vs_SPs} shows the
number of SPs as a function of $N$. The SPs for different models
with similar sizes have been studied using the gradient minimization
and Newton-Raphson's methods, though the final number of SPs is always open to debate \cite{2002JChPh.116.3777D,2003JChPh.11912409W},
whereas using the NPHC method we can find all the SPs with confidence.
In the gradient minimization method, it is difficult to obtain the
SPs with higher indices \cite{2002JChPh.116.3777D}. In the NPHC, however, 
all the SPs are treated equally irrespective of
their indices. The right panel in Figure \ref{fig:N_vs_SPs} shows
the index $I/(N-1)$ vs number of SPs for a given $N$. Apparently,
the number of minima grows linearly in this model, which is contrary
to the conventional wisdom \cite{2003JChPh.11912409W,2002JChPh.116.3777D}.
However, the number of SPs indeed grows exponentially with increasing
system size as expected \cite{2003JChPh.11912409W,2002JChPh.116.3777D}.
We also observe that the global minimum is always the configuration
with all $\theta_{i}=0,i=1,\dots,N-1$, and that at the global minimum,
the energy density $V/N$ is zero. In Figure \ref{fig:energy_density_vs_index},
we plot $V/N$ vs $I$ for $N=7,9,11,13$. The apparent non-linear
relation between these two quantities in the plots show a different
behaviour to the linear relationship observed for the Lennard-Jones models.  

\begin{figure}
\includegraphics[width=4cm,height=3cm]{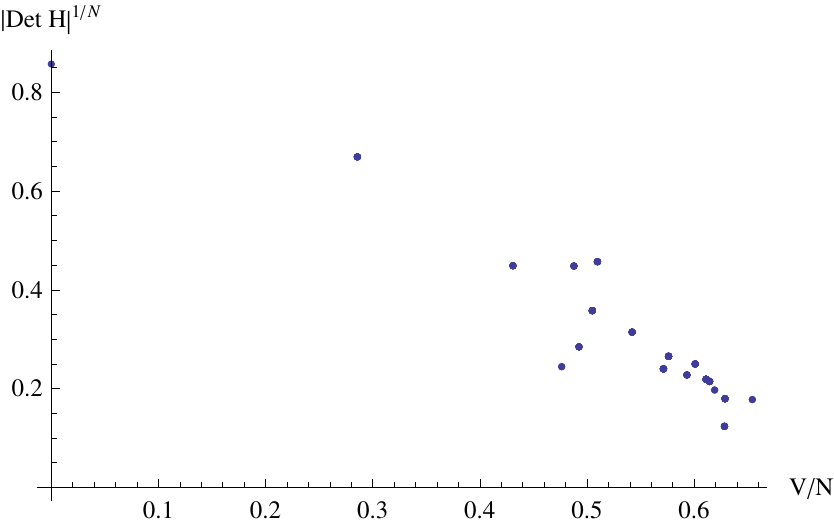}\includegraphics[width=4cm,height=3cm]{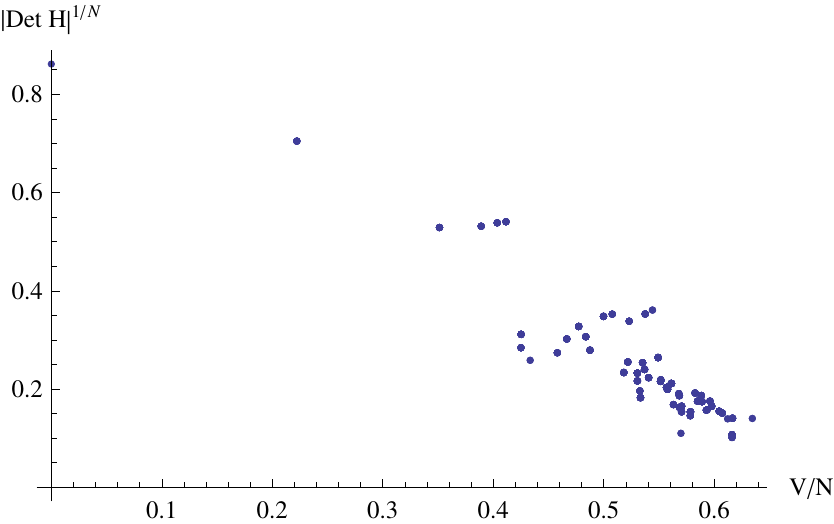}

\includegraphics[width=4cm,height=3cm]{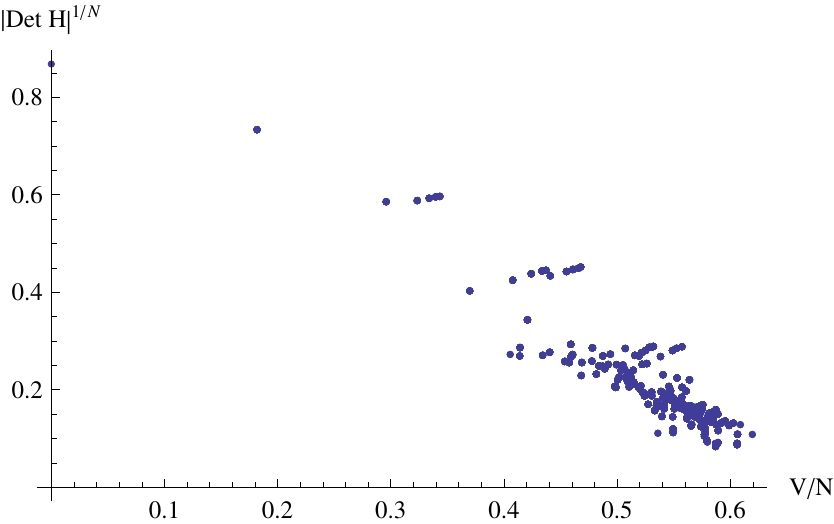}\includegraphics[width=4cm,height=3cm]{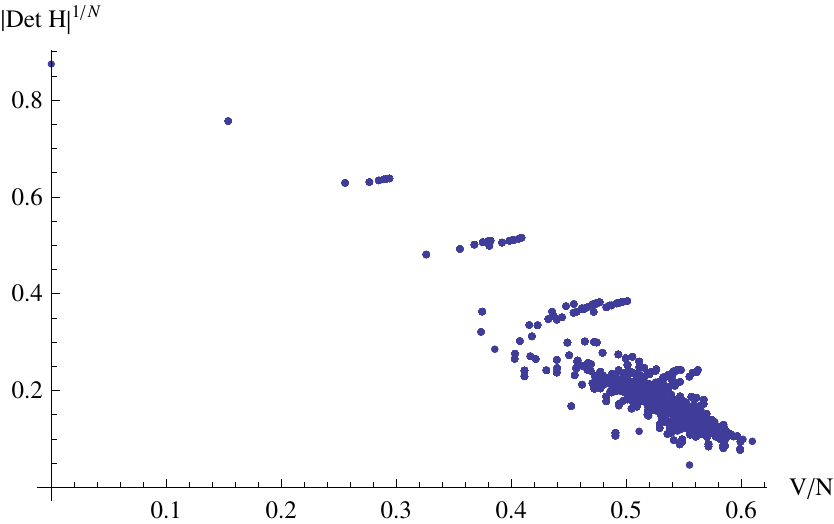}

\caption{Plot of $V/N$ vs $|\mbox{Det }\mathcal{H}|^{1/N}$ starting from
upper left corner for $N=7,9,11,13$, respectively. Though these small
sized systems can in no way be considered as representatives of the
model in the thermodynamic limit, we still see the expected singularity
of $|\mbox{Det }\mathcal{H}|^{1/N}$ at $V/N\rightarrow0.5$. \label{fig:energy_density_vs_det}}
\vspace{-0.7cm}
\end{figure}

Although it is quite difficult to claim any result in the thermodynamic
limit using such small size systems, we note that the minimum value
of the $|\mbox{Det}\mathcal{H}|^{1/N}$ is always either at the SPs
for which all $\theta_{i}\in\{0,\pi\}$ for all $i=1,\dots,N-1$,
or for the SPs for which $\theta_{i+1}-\theta_{i}=\frac{2\pi k}{N}\mbox{ mod }2\pi$
for all $i=1,\dots,N$, with $k$ being an integer. Using a recently spelled out condition for
a class of spin-glass models to have phase transition, that $|\mbox{Det }\mathcal{H}|^{1/N}$
evaluated at a class of stationary points tend to go to zero at the
critical energy in the $N\rightarrow\infty$ limit, the phase transition
in our model would occur at this specific class of solutions \cite{PhysRevE.80.060103}.
For $N=7,9,11,13$, the corresponding plots are drawn in Figure \ref{fig:energy_density_vs_det}
where the plot has started being filled up by densely populated points
around the critical energy density $V/N\sim0.5$. Hence, we have reproduced
the results that Kastner worked out from other approach in \cite{2010arXiv1011.5050K}
by just studying these small size systems. These particular SPs are usually neither minima nor maxima.

It would be interesting to get all the SPs for $\alpha\in(1,2]$,
find out the relevant SPs for the phase transition and extrapolate
the analysis in the thermodynamic limit as suggested in \cite{2010arXiv1011.5050K}.
Verifying a recent conjecture that the critical energy density of
a class of spin glass models can be computed using the Ising-like
SPs only (see\cite{2011PhRvL.106e7208C} for the definition of these
SPs and the conjecture), will also be another important application
of the NPHC method.

In this Letter, we have described and applied a novel method, called
the numerical homotopy continuation (NPHC) method, that finds \textit{all}
SPs of a given potential, provided that the potential has polynomial-like
non-linearity. As shown in this Letter, even if the given potential
is not apparently in the polynomial form, its stationary equations
could be transformed into a polynomial form by adding suitable constraint
equations. So the method is very widely applicable. The ability of
finding all SPs and it being completely parallelizable makes the
method quite promising to study many areas of theoretical physics
and chemistry, for example, finding all the SPs and hence all the
minima of the Lennard-Jones potential and its numerous variants; to
obtain the string vacua of the models for which the symbolic algebraic
geometry methods fail due to their algorithmic complexities; to study
phase transitions in various spin glass models with the above mentioned
criterion on the hessian determinant etc. We anticipate that this
work will give a thrust to the current research in the related areas.
\begin{acknowledgments}
DM was supported by the U.S. Department of Energy grant under contract
no. DE-FG02-85ER40237 and Science Foundation Ireland grant 08/RFP/PHY1462. 
\end{acknowledgments}

\vspace{-0.5cm}

\bibliographystyle{apsrev4-1}
\addcontentsline{toc}{section}{\refname}\bibliography{bibliography}

\end{document}